\documentclass[11pt]{article} 
\usepackage[english]{babel}
\usepackage{mathtools}
\usepackage{amsmath} 
\usepackage{comment}

\usepackage{setspace}

\usepackage[font={it}]{caption}
\usepackage{amssymb}  
\usepackage{makeidx} 
\usepackage{graphicx} 
\usepackage{color}
\usepackage[dvipsnames]{xcolor}
\usepackage{epstopdf}

\usepackage{fullpage}
\usepackage{enumitem}
\usepackage[section]{placeins} 
\usepackage{flafter}
\usepackage{geometry}

\usepackage{subcaption}
\usepackage{wrapfig}

\usepackage{hyperref} 

\usepackage{makeidx}
\makeindex

\usepackage{tabularx}
\usepackage{multicol}
\usepackage{makecell}

\usepackage{framed}
\usepackage{appendix}
\usepackage[nottoc]{tocbibind}
\usepackage{fancyhdr}

\usepackage{csquotes}

\usepackage{emptypage}
\usepackage[round]{natbib}
\usepackage{comment}
\usepackage{fancyhdr}
\usepackage{fancyvrb}
\usepackage{comment}
\usepackage{bbm}
\usepackage{booktabs}

\usepackage{longtable}
\usepackage{arydshln}

\usepackage{multirow}
\usepackage{fancybox} 
\usepackage{tikz} 
\usetikzlibrary{arrows.meta, graphs}

\usepackage{pdflscape} 

\usepackage{authblk} 

\everymath=\expandafter{\the\everymath\displaystyle}

\usepackage{epstopdf} 

\makeatletter 
\renewcommand\@biblabel[1]{#1} 
\makeatother %

\title{Optimal Scaling transformations to model non-linear relations in GLMs with ordered and unordered predictors}

\author[1]{S.J.W. Willems \footnote{Corresponding author; s.j.w.willems@fsw.leidenuniv.nl}}
\author[2]{A.J. Van der Kooij}
\author[2,3]{J.J. Meulman}

\affil[1]{Methodology \& Statistics, Institute of Psychology, Leiden University, Leiden, Netherlands}
\affil[2]{LUXs data science BV, Leiden, Netherlands}
\affil[3]{Department of Statistics, Stanford University, Stanford, California CA 94305}

\date{Version: September 1, 2023}

\begin{document}


\newcommand{\betaB}{\boldsymbol{\beta}}
\newcommand{\XB}{\mathbf{X}}
\newcommand{\xB}{\mathbf{x}}
\newcommand{\etaB}{\boldsymbol{\eta}}
\newcommand{\XBidot}{\XB_{i \ast}}
\newcommand{\xBidot}{\xB_{i \ast}}
\newcommand{\phiB}{\boldsymbol{\varphi}}
\newcommand{\XBdotk}{\XB_{\ast k}}
\newcommand{\xBdotk}{\xB_{\ast k}}

\newcommand{\XBdotl}{\XB_{\ast l}}
\newcommand{\XBdotp}{\XB_{\ast p}}
\newcommand{\XBdotONE}{\XB_{\ast1}}

\newcommand{\zB}{\mathbf{z}}

\newcommand{\yB}{\mathbf{y}}
\newcommand{\uB}{\mathbf{u}}

\newcommand{\phiXB}{\varphi(\XB)}
\newcommand{\phiBXB}{\boldsymbol{\varphi}(\XB)}
\newcommand{\phiXBi}{\varphi(\XB_i)}
\newcommand{\phikXk}{\varphi_k(\xB_k)}
\newcommand{\phikXik}{\varphi_k(x_{ik})}
\newcommand{\phiXBdotk}{\varphi_k(\XBdotk)}
\newcommand{\phixBdotk}{\varphi_k(\xBdotk)}
\newcommand{\phiXBdotl}{\varphi_l(\XBdotl)}
\newcommand{\phixBdotl}{\varphi_l(\xBdotl)}

\newcommand{\phiXBidot}{\varphi(\XBidot)}
\newcommand{\phiBXBidot}{\boldsymbol{\varphi}(\XBidot)}

\newcommand{\phikxik}{\varphi_k(x_{ik})}
\newcommand{\phixB}{\varphi(\xB)}
\newcommand{\phixBi}{\varphi(\xB_i)}
\newcommand{\philxil}{\varphi_l(x_{il})}

\newcommand{\GB}{\mathbf{G}}
\newcommand{\vB}{\mathbf{v}}
\newcommand{\DB}{\boldsymbol{D}}

\newcommand{\lnorm}{\left\Vert}
\newcommand{\rnorm}{\right\Vert}

\newcommand{\tildebetaB}{\widetilde{\betaB}}
\newcommand{\tildeetaB}{\widetilde{\boldsymbol{\eta}}}

\newcommand{\OmegaI}{\mathit{\Omega}}

\newcommand{\OmegaB}{{\color{red}\Omega}}

\newcommand{\GradientLBeta}{\boldsymbol{\nabla}_{l} (\betaB)}
\newcommand{\GradientLHatBeta}{\boldsymbol{\nabla}_{l} (\widetilde{\betaB})}
\newcommand{\HessianLBeta}{\mathbf{H}_{l} (\betaB)}
\newcommand{\HessianLHatBeta}{\mathbf{H}_{l} (\widetilde{\betaB})}
\newcommand{\HessianInvLBeta}{\mathbf{H}_{l}^{-1} (\betaB)}
\newcommand{\HessianInvLHatBeta}{\mathbf{H}_{l}^{-1} (\widetilde{\betaB})}

\newcommand{\GradientLVk}{\boldsymbol{\nabla}_{l} (\vB_k)}
\newcommand{\GradientLHatVk}{\boldsymbol{\nabla}_{l} (\widetilde{\vB}_k)}
\newcommand{\HessianLVk}{\mathbf{H}_{l} (\vB_k)}
\newcommand{\HessianLHatVk}{\mathbf{H}_{l} (\widetilde{\vB}_k)}
\newcommand{\HessianInvLVk}{\mathbf{H}_{l}^{-1} (\vB_k)}
\newcommand{\HessianInvLHatVk}{\mathbf{H}_{l}^{-1} (\widetilde{\vB}_k)}

\newcommand{\GradientLBetak}{\boldsymbol{\nabla}_{l} (\beta_k)}
\newcommand{\GradientLHatBetak}{\boldsymbol{\nabla}_{l} (\widetilde{\beta}_k)}
\newcommand{\HessianLBetak}{\mathbf{H}_{l} (\beta_k)}
\newcommand{\HessianLHatBetak}{\mathbf{H}_{l} (\widetilde{\beta}_k)}
\newcommand{\HessianInvLBetak}{\mathbf{H}_{l}^{-1} (\beta_k)}
\newcommand{\HessianInvLHatBetak}{\mathbf{H}_{l}^{-1} (\widetilde{\beta}_k)}

\newcommand{\GradientLEta}{\boldsymbol{\nabla}_{l} (\etaB)}
\newcommand{\GradientLHatEta}{\boldsymbol{\nabla}_{l} (\widetilde{\etaB})}
\newcommand{\HessianLEta}{\mathbf{H}_{l} (\etaB)}
\newcommand{\HessianLHatEta}{\mathbf{H}_{l} (\widetilde{\etaB})}
\newcommand{\HessianInvLEta}{\mathbf{H}_{l}^{-1} (\etaB)}
\newcommand{\HessianInvLHatEta}{\mathbf{H}_{l}^{-1} (\widetilde{\etaB})}

\newcommand{\GradientNegLHatEta}{\boldsymbol{\nabla}_{-l} (\widetilde{\etaB})}
\newcommand{\HessianNegLHatEta}{\mathbf{H}_{-l} (\widetilde{\etaB})}
\newcommand{\HessianInvNegLHatEta}{\mathbf{H}_{-l}^{-1} (\widetilde{\etaB})}

\maketitle

\begin{abstract}
\noindent In Generalized Linear Models (GLMs) it is assumed that there is a linear effect of the predictor variables on the outcome. However, this assumption is often too strict, because in many applications predictors have a nonlinear relation with the outcome. Optimal Scaling (OS) transformations combined with GLMs can deal with this type of relations. 	
Transformations of the predictors have been integrated in GLMs before, e.g.\ in Generalized Additive Models. However, the OS methodology has several benefits. 
For example, the levels of categorical predictors are quantified directly, such that they can be included in the model without defining dummy variables. This approach enhances the interpretation and visualization of the effect of different levels on the outcome. 
Furthermore, monotonicity restrictions can be applied to the OS transformations such that the original ordering of the category values is preserved. This improves the interpretation of the effect and may prevent overfitting. 
The scaling level can be chosen for each individual predictor such that models can include mixed scaling levels. In this way, a suitable transformation can be found for each predictor in the model. 
The implementation of OS in logistic regression is demonstrated using three datasets that contain a binary outcome variable and a set of categorical and/or continuous predictor variables.

\end{abstract}

\newpage

\section{Introduction}

Linear models are often used to model relations between a numeric outcome variable and a set of predictor variables. The ordinary least squares regression model (OLS) assumes normally distributed errors and linearity in the predictors. Due to these assumptions, the application to real data is sometimes limited. 
For example, consider a medical application in which the relation between the binary outcome of getting a particular disease and the predictor variable age is modeled. First of all, the binary outcome cannot be modeled with the standard linear regression model due to the assumption of normally distributed errors. Furthermore, due to their weaker immune systems, it may be expected that both young children and elderly people are more susceptible to the disease than people of intermediate ages. In such situations, the relation between age and the probability of getting the disease will have an u-shape and thus the linearity assumption is too strict. Hence, for these types of situations, the ordinary linear model is not appropriate.

To increase the applicability of the linear model, several extensions have been developed. 

One extension is to allow for a nonlinear relation between the linear combination of the predictor variables and the outcome via a link function. This type of models are known as Generalized Linear Models (GLMs, \citet{Book:GLMsMcCullaghNelder}). GLMs do not assume normally distributed errors and are therefore applicable if errors are distributed differently. A frequently used GLM for binary outcomes is the logistic regression model, which uses the logit link function to transform the linear predictor into the unit interval to model probabilities.

A second extension is by transforming the variables. This is done in, for example, additive models (\citet{Art:AdditiveModels_Backfitting_FriedmanStuetzle, Book:GeneralizedAdditiveModels_HastiTib, Art:WinsbergRamsayMonotonicTransformationsToAdditivity}) and Optimal Scaling regression (OS-regression) \citep{Art:OSinRegression, book:Gifi, Art:SPSSCategoriesCATREG}. The predictor variables are transformed using either a parametric or a nonparametric function. 

In this paper, we will integrate two extensions of ordinary linear models by combining GLMs with optimal scaling techniques. As a result, a nonlinear link function (as in a GLM) is used to model the relation between the response variable and a linear combination of \emph{transformed} predictor variables (as in the OS approach). Hence, the important difference between a regular GLM and a GLM with OS lies in the transformation of the predictor variables. We will first explain the OS algorithm for ordinary linear models, before we show how OS can be integrated in the Newton-Raphson method which is used to fit a GLM model.

\vspace{\baselineskip}
Initially, OS was developed to transform nominal or ordinal categorical variables into quantitative data by finding optimal numeric values for the category values. This process was referred to as \emph{quantifying qualitative data} by \citet{Art:QuantitativeAnalysisofQualitativeData} and the resulting transformations are called \emph{quantifications}, denoted as $\varphi_k(\xB_k)$ for variable $k$. 

The quantifications can also be written in matrix form as $\varphi_k(\xB_k) = \GB_k \vB_k$. Here, $\vB_k$ is a vector with the quantifications for each category of variable $k$, and $\GB_k$ is an indicator matrix that represents the observed category values in $\xB_k$. Namely, the number of columns in this matrix is equal to the number of categories and each row contains only zero's and a single one where the one is placed in the column that corresponds with $i$'s observed category. 

Although the OS methodology was originally developed for categorical data, it can also be applied to non linearly transform numeric data. In this case, each unique observation of the numeric variable is interpreted as an individual category and they are modeled in the same way as for categorical predictors. Hence, if all objects have unique values, $\GB_k$ is a permuted identity matrix.

In OS-regression, the response $y_i$ of observation $i$ is modeled as a linear combination of the quantifications of the $p$ observed predictors. Hence, the model is as follows
\begin{displaymath}
\textstyle{y_i = \sum_{k=1}^{p} \beta_k \phikXik + \epsilon_i},
\end{displaymath} 
where $\epsilon_i$ is the error term.

\vspace{\baselineskip}
The type of transformation (also called \emph{scaling level} in the categorical data analysis context) is chosen for each individual variable and may thus differ among predictors. The combination of coefficients and transformations calculated by the algorithm, optimally describe the relation between the response and the predictors under the restrictions set by the chosen scaling levels. Several types of scaling levels can be chosen.

Usually a step-function is chosen for categorical predictors with few levels which can either be monotone or nonmonotone, depending on whether the ordering of the levels should be preserved. \citet{Art:KruskalMonotonicTransformations} described one of the first algorithms to find monotonic step transformations in multidimensional scaling and a similar technique is applied in OS. The \textit{nonmonotone step-function} and \textit{monotone step-function} scaling levels are often referred to as respectively the \textit{nominal} and \textit{ordinal} scaling levels, as they mostly resemble the characteristics of nominal and ordinal categorical variables.

If the predictor has many levels (e.g.\ for a numeric variable), some smoothing may be appropriate to avoid overfitting and to improve interpretation. In these cases, either a \textit{monotone spline function} or \textit{nonmonotone spline function} can be fit, again depending on whether the ordering of the categories should be preserved. I-splines (as described by \citet{Art:RamsayMonotonicSplines}) are used to fit the (non)monotonic spline function. 


The nonmonotone functions, especially the step-function, are the least restrictive scaling levels as the monotonicity assumption in the other scaling levels adds more restrictions. The scaling level for each predictor is usually chosen depending on the expected relation between that predictor and the outcome, or based on the transformation results of a preliminary analysis with the least restrictive scaling levels showing the shapes of the relations. 

Additionally, a \textit{numeric} scaling level can chosen for a predictor, for example if either the relation is a priori assumed to be linear, or if a non-linear scaling level was chosen and the resulting transformation turns out being (close to) linear. Note that if a numeric scaling level is chosen for all predictors, the GLM-OS will give the same output as a ordinary GLM.

\vspace{\baselineskip}	
In this paper, we will describe how OS can be integrated in the Newton-Raphson algorithm to find optimal quantifications for the predictor variables in a GLM. This combination of methods is called the Generalized Linear Model with Optimal Scaling (GLM-OS) or the generalized version of Optimal Scaling regression (GOS-regression). Since the OS method allows for nonlinear transformations, the term \emph{linear predictor} that is used for the linear combination of transformed predictors in the GLM literature may be confusing. Therefore, it is referred to in this paper as the \emph{weighted sum} or \emph{linear combination} of transformed predictors. 

Although applicable to more GLMs, we will focus on logistic regression with OS transformations and apply this model to three datasets. Each of these datasets has different types of predictor variables, which allows us to illustrate the benefits of OS with respect to visualization, interpretation, and predictability.


\section{Optimal scaling in linear regression}
\label{GLM-OS_sec:introOSforLM} 
In this section, we will explain how optimal scaling transformations are integrated in linear regression. To keep this explanation concise, we only show the basics and leave out the details and extensions. For more details about the OS-regression algorithm, including some adjustments to optimize calculation time, we refer to \citet{proefschrift:AnitaKooij} and \citet{Art:MeulmanVanDerKooijROSS}.

\subsection{Model and notation}
Let $\XB$ be the data matrix of dimension $n \times p$ where $n$ and $p$ are the number of objects and predictors respectively. 
The $n$ observed values of the response variable are collected in the vector $\yB$.

In ordinary least squares regression (OLS), the outcome is modeled as a linear combination of the predictors, i.e.\ $\textstyle{y_i = \sum_{k=1}^{p} \beta_k  x_{ik} + \epsilon_i}$, where $\epsilon_i$ is the error term. In the optimal scaling setting, the original observed values $\xB_k$ of each predictor variable $k$, for $k = 1, \ldots, p$, are transformed and replaced by their quantifications that is denoted as $\phikXk$. The outcome $\yB$ is assumed to be centered and therefore no intercept is required. Hence, the OS-regression model is $\textstyle{\yB = \sum_{k=1}^{p} \beta_k \; \phikXk + \boldsymbol{\epsilon}}$.

The quantifications for all $n$ observations can be written in matrix form. Let $C_k$ be the number of unique observed values for predictor $k$, and denote by $\GB_k$ the indicator matrix of dimensions $n \times C_k$. Each $i$th row of $\GB_k$ consists of $C_k-1$ zero's and a single one, placed in column $x_{ik}$, where $\xB_k$ is assumed to consist of either consecutive category numbers from 1 to $C_k$, or the rank numbers of unique values for numeric predictors. Furthermore, let $\vB_k$ be the $C_k \times 1$ vector that contains the $C_k$ quantifications of predictor $k$. Then, $\GB_k \vB_k$ is the $n$-vector of the transformed value for each object, i.e.\ $\phikXk = \GB_k \vB_k$. Using this notation, the linear regression model with optimal scaling quantifications in matrix form can be written as
\begin{equation}
\label{GLM-OS_eq:OS-regressionmodel}
\textstyle{
	\yB = \sum_{k=1}^{p} \beta_k \;  \phikXk + \boldsymbol{\epsilon} = \sum_{k=1}^{p} \beta_k \;  \GB_k \vB_k + \boldsymbol{\epsilon} .
}
\end{equation}
The matrices $\GB_1, \ldots, \GB_p$ are derived from the data, and coefficients $\beta_1, \dots, \beta_p$ and quantifications $\vB_1, \ldots, \vB_p$ need to be estimated.

\subsection{Model estimation}
The loss function corresponding to the OS-regression model in \eqref{GLM-OS_eq:OS-regressionmodel} is written as
\begin{equation}
\label{GLM-OS_eq:LossFunctionOS-regression_MatrixNotation}
\textstyle{
	L(\vB_1, \ldots, \vB_p; \beta_1, \ldots, \beta_p) = \lnorm \yB - \sum_{k=1}^{p} \beta_k \GB_k \vB_k \rnorm^2.
}
\end{equation}
To fit the model, the loss function should be minimized over both the model coefficients $\beta_1, \ldots, \beta_p$, and the quantifications $\vB_1, \ldots, \vB_p$ simultaneously, where the quantifications are restricted according to their scaling level, as described above these are nominal and ordinal step or spline functions.
As an infinite number of combinations of model coefficients and quantifications will optimize this function, the latter are standardized to ensure a unique solution. 

Since no closed-form solution is available to minimize loss function \eqref{GLM-OS_eq:LossFunctionOS-regression_MatrixNotation} over all parameters simultaneously, the quantifications and coefficients are optimized iteratively for one variable at a time, until convergence. This type of algorithm is referred to as \textit{alternating least squares} in the psychometric literature \citep{book:Gifi, Art:OSinRegression}, since the least squares solution is calculated by alternating the estimation of optimal quantifications and model coefficients for one variable at a time. In the statistical literature it is called \textit{backfitting} and has been extensively used to fit Additive Models and GAMs \citep{Art:AdditiveModels_Backfitting_FriedmanStuetzle, Book:GeneralizedAdditiveModels_HastiTib}. A variety of other terms is present in the literature, like \textit{component-wise update} and \textit{block relaxation}, but it is currently usually referred to as \textit{coordinate descent}.

In the initialization step, standardized values of the observed variables are used as starting values for the quantifications $\vB_1, \ldots, \vB_p$, and the ordinary least squares coefficients for these standardized quantifications are used as starting values for $\beta_1, \ldots, \beta_p$. If a numeric scaling level is chosen for variable $k$, the only transformation is the conversion to $Z$-scores.

After initialization, the parameters are updated for a single variable at the time. At each iteration, all regression coefficients and variables are assumed to be fixed, except for the variable $k$ that is currently (conditionally) optimized. All fixed terms are merged into a single vector denoted by $\uB_k$ and variable $k$ is then separated from this fixed part, i.e.\
\begin{equation}
\label{GLM-OS_eq:LossFunctionOS-regression_MatrixNotation_withUk}
\textstyle{
	L(\vB_k, \beta_k) = \lnorm \yB - \sum_{l\neq k} \beta_l \GB_l \vB_l -  \beta_k \GB_k \vB_k \rnorm^2 = \lnorm \uB_k -  \beta_k \GB_k \vB_k \rnorm^2.
}
\end{equation}

If variable $k$'s scaling level is not numeric, quantifications $\vB_k$ need to be updated. While updating $\vB_k$, it is assumed that $\beta_k$ is fixed, which enables us to calculate the solution for $\vB_k$ as the ordinary least squared solution for \eqref{GLM-OS_eq:LossFunctionOS-regression_MatrixNotation_withUk} with respect to $\vB_k$. Hence, if $\widetilde{\vB}_k$ is the current estimate of $\vB_k$, then it is updated as 
\begin{align}
\label{eq:UnrestrictedVUpdate_OrdinaryOSRegr}
\widetilde{\vB}_k^+ 
&= \left\{ ( \widetilde{\beta}_k \GB_k )^T  \widetilde{\beta}_k \GB_k \right\}^{-1} ( \widetilde{\beta}_k \GB_k )^T \uB_k \nonumber\\
&= \left\{ \widetilde{\beta}_k^2  \GB_k^T \GB_k \right\}^{-1}  \GB_k ^T \widetilde{\beta}_k^T \uB_k \nonumber \\
&= \widetilde{\beta}_k^{-1} \DB_k^{-1} \GB_k ^T \uB_k, 
\end{align} 
where $\widetilde{\beta}_k$ is the current estimate of $\beta_k$ and $\DB_k  = \GB_k^T \GB_k$. Actually, since $\widetilde{\vB}_k$ will be standardized later and $\widetilde{\beta}_k^{-1}$ is only a scalar, it can be dropped, retaining only its sign.

This result $\widetilde{\vB}_k^+$ is actually the solution to the optimal scaling problem with least restrictive scaling level, namely the nonmonotone step-function. For the other scaling levels, restrictions have to be applied to $\widetilde{\vB}_k^+$. For the ordinal scaling level, weighted monotonic regression (\citet{Art:KruskalMonotonicTransformations}) is applied, resulting in a monotonic step function. For the nonmonotone and monotone spline restrictions (with a specified number of knots and degree of the polynomial functions, \citet{Art:RamsayMonotonicSplines}) are fitted to the unrestricted solution. After the appropriate restrictions have been applied, the result is standardized to ensure a unique solution. This restricted and standardized solution is then the current estimate $\widetilde{\vB}_k$ of $\vB_k$. 

Once the quantifications of the $k$th variable have been updated, model parameter $\beta_k$ is estimated by again using the ordinary least squares solution for loss function \eqref{GLM-OS_eq:LossFunctionOS-regression_MatrixNotation_withUk} in which $\GB_k \vB_k$ is now fixed. Hence, the updated value for $\beta_k$ is calculated as
\begin{align}
\widetilde{\beta}_k^+ 
&= \left\{ ( \GB_k \widetilde{\vB}_k )^T \GB_k \widetilde{\vB}_k \right\}^{-1} (\GB_k \widetilde{\vB}_k )^T \uB_k \nonumber\\
&= \left\{ \widetilde{\vB}_k^T  \DB_k  \widetilde{\vB}_k \right\}^{-1} \widetilde{\vB}_k^T  \GB_k^T \uB_k.  
\end{align}

The algorithm continues updating the quantifications and model parameters for the other variables. This process continues until the loss measured by \eqref{GLM-OS_eq:LossFunctionOS-regression_MatrixNotation} does not change anymore. 

The final estimates of the model coefficients and quantifications (denoted as $\widehat{\beta}_1, \ldots, \widehat{\beta}_p$ and $\widehat{\vB}_1, \ldots, \widehat{\vB}_p$) are the updates from the last iteration. Usually the final estimates of the quantifications ($\widehat{\vB}_k$) are plotted against the original values of variable $k$ to visualize the transformations. 

Note that $\GB_k$ and $\DB_k$ are sparse but potentially big matrices which can make the above calculations very inefficient. Therefore, in the implementation of the algorithm, advantage is taken of the sparse structure of these matrices, only using the diagonal of $\DB_k$.

\begin{framed}
	\noindent \textbf{OS-regression algorithm:}
	\begin{description}
		\item[Initialization:] Create $\GB_1, \ldots, \GB_p$ based on the data, and initialize the model parameters $\widetilde{\beta}_1, \ldots, \widetilde{\beta}_p$ and $\widetilde{\vB}_1, \ldots, \widetilde{\vB}_p$.
		
		\item[Cycle:] For $k=1, \ldots, p$, do:
		\begin{description}
			\item[Step 1:] Calculate $\uB_k = \yB - \sum_{l\neq k} \beta_l \GB_l \vB_l $. 
			
			\item[Step 2:] If the scaling level of variable $k$ is not numeric, calculate the unrestricted estimates of the quantifications of $k$ as 
			\begin{displaymath}
			\widetilde{\vB}_k^{+} =  \widetilde{\beta}_k^{-1} \DB_k^{-1} \GB_k ^T \uB_k.
			\end{displaymath}
			
			Apply appropriate scaling restrictions to $\widetilde{\vB}_k^{+}$ and standardize the result. 
			
			\item[Step 3:] Update the estimate for model coefficient $\beta_k$ as 
			\begin{displaymath}
			\widetilde{\beta}_k^{+} = \left\{ \widetilde{\vB}_k^T  \DB_k  \widetilde{\vB}_k \right\}^{-1} \widetilde{\vB}_k^T  \GB_k^T \uB_k.  
			\end{displaymath}
			
		\end{description}
		\item[Convergence:] Repeat the cycle until convergence criteria are met.

	\end{description}
\end{framed}


\section{Optimal scaling in generalized linear models}

In this section we will explain how the OS procedure can be integrated in the Newton-Raphson algorithm used to fit GLMs. After describing the Newton-Raphson algorithm as it is used to fit regular GLMs, we will show how it can be modified to include OS transformations. Then we will show the specific example of how optimal scaling transformations can be calculated for the logistic regression model. This model will also be used for the data illustrations in the next section. 

\subsection{GLM-OS model and notation}
\label{GLM-OS_subsec:GLM-OS_ModelAndNotation}

For GLM-OS we use notation that is similar to the notation used for OS-regression. Hence, let $\mathbf{X}$ and $\yB$ again be the data matrix and the vector with the outcome. In a GLM, the outcome is not centered and thus these models include an intercept. In the GLM-OS setting, we therefore incorporate the intercept as the regression coefficient $\beta_0$ multiplied by a vector of ones, which is denoted as $\xB_0$ and included in the data matrix $\mathbf{X}$.\\

\noindent A GLM consists of three components, namely
\begin{description}
	\item[1)]a random component that specifies the distribution of the response variable given the predictors;
	\item[2)]a linear combination of the predictor variables, denoted as $\etaB= \beta_0 \xB_0 + \beta_1 \xB_1 + \ldots + \beta_p \xB_p$;
	\item[3)]an invertible link function $g$ which models the relation between the linear combination of object $i$ and $i$'s response $y_i$, i.e.\ $\textstyle{g(\mu_i ) = \eta_i = \beta_0 x_{i0} + \ldots + \beta_p x_{ip}}$ where $\textstyle{\mu_i = E(Y_i)}$.
\end{description}

\vspace{\baselineskip}
To extend GLMs to include optimal scaling transformation the linear combination of predictors is replaced by a linear combination of the quantifications, so
\begin{equation}
\label{GLM-OS_eq:GLM-OSLinearPredictor}
\textstyle{
	\etaB = \sum_{k=0}^p \beta_k \varphi_k(\xB_k) = \sum_{k=0}^p \beta_k  \GB_k \vB_k.
}
\end{equation}
To fit the GLM-OS, coefficients $\beta_0, \dots, \beta_p$ and quantifications $\vB_1, \ldots, \vB_p$ need to be estimated. Note that, to represent the intercept, $\varphi_0(\xB_0) = \boldsymbol{1}_n$, and consequently $\GB_0 = \boldsymbol{1}_n$ and $\vB_0 = \{1\}$, are fixed, and hence these terms do not have to be estimated in each iteration.

\subsection{Model estimation}

The maximum likelihood approach is used to estimate GLMs. The exact form of the likelihood function depends on the random component of the GLM and the link function. The log-likelihood is is a function of the linear combination of predictors and is denoted as $l(\etaB)$. In a GLM $\etaB$ only depends on parameters $\beta_0, \ldots, \beta_p$, while in GLM-OS it depends on both $\beta_0, \ldots \beta_p$ and $\vB_1, \ldots \vB_p$. 

There is no closed-form solution to maximize the (log-)likelihood functions, hence a numerical method is required to find the maximum likelihood estimator (MLE). For GLMs, usually the Newton-Raphson method is used.

\subsubsection{Newton-Raphson method for GLMs}
\label{GLM-OS_subsec:NewtonRaphsonForOrdinaryGLM} 

The GLM fitting algorithm aims to find the roots of the gradient by using the Newton-Raphson algorithm. This method iteratively improves the initial starting values via the first-order Taylor approximation of the gradient $\GradientLBeta$ of the log-likelihood around the current guess $\widetilde{\betaB}$. Hence, the solutions are found as follows
\begin{align*}
\mathbf{0}  = \GradientLBeta &  \approx \GradientLHatBeta + \HessianLHatBeta (\betaB - \widetilde{\betaB}) \\
-  \HessianLHatBeta \; \betaB &  \approx \GradientLHatBeta - \HessianLHatBeta \; \widetilde{\betaB}  \\
\betaB  &  \approx \widetilde{\betaB} - \HessianInvLHatBeta \; \GradientLHatBeta,
\end{align*}
where 
\begin{itemize}
	\item $\widetilde{\betaB} $ is the current estimate of $\betaB$;
	\item $\HessianLHatBeta$ is the Hessian matrix containing all the second-order partial derivatives of $l(\etaB)$ w.r.t.\ $\betaB$ evaluated at $\widetilde{\betaB} $;
	\item $\GradientLHatBeta$ is the gradient vector that contains the first-order partial derivatives of $l(\etaB)$ w.r.t.\ $\betaB$ evaluated at $\widetilde{\betaB} $.
\end{itemize}
Then, the current estimate $\widetilde{\betaB}$ is updated in each iteration as
\begin{equation}
\widetilde{\betaB}^{+} = \widetilde{\betaB} - \HessianInvLHatBeta \; \GradientLHatBeta.
\end{equation}

Each update $\textstyle{\widetilde{\betaB}^{+}}$ should be a better approximation of the root than the previous estimate $\widetilde{\betaB}$ and the algorithm repeatedly updates these estimates until the convergence criteria are met.

In some applications an approximation of $\HessianInvLHatBeta$ is used to simplify the calculations. For example, in Fisher's Scoring method, the Hessian is replaced by its expectation. In some cases, the Hessian and its expectation are identical, in which case Newton-Raphson and Fisher's scoring method are equivalent. If $\HessianLHatEta$ is not  a diagonal matrix, it can be approximated by a diagonal matrix to reduce calculation time. For example, \citet{Art:SimonsEtAlLeastSquares} and \citet{Art:OSCox_Willems} fitted Cox' Proportional Hazards model in the context of regularization and OS transformations respectively, and approximated the full Hessian matrix by its diagonal.

If it is easier to do calculations with the negative log-likelihood, the algorithm is modified such that it finds the minimum of $-l(\etaB)$. In this case the algorithm does essentially not change except that it now uses the the gradient and Hessian of the negative log-likelihood $-l(\etaB)$ to repeatedly update $\widetilde{\betaB}$ to find $\widehat{\betaB}$.

\subsubsection{Modification of the Newton-Raphson method to fit GLM-OS}
\label{GLM-OS_subsec:ModifiedNewtonRaphsonScoringMethod}

To estimate the GLM-OS model the coefficients $\beta_1, \dots, \beta_p$ and quantifications $\vB_1, \ldots, \vB_p$ that maximize the log-likelihood function need to be computed. Hence, the following equations need to be solved,
\begin{displaymath}
\boldsymbol{\nabla}_{l} (\vB_1) = \ldots = \boldsymbol{\nabla}_{l} (\vB_p) =\mathbf{0},
\end{displaymath}
and
\begin{displaymath}
\boldsymbol{\nabla}_{l} (\beta_0) = \ldots = \boldsymbol{\nabla}_{l} (\beta_p) = 0.
\end{displaymath}

Since there is no closed-form solution to derive the parameters simultaneously, they will be calculated iteratively for one variable $k$ at the time, as was done in the OS-regression algorithm.
After initialization, the algorithm iterates over all $k = 0, \ldots, p$ predictors and updates first the quantifications $\vB_k$ (unless $k = 0$) and then model coefficient $\beta_k$. 

As when updating $\betaB$ in ordinary GLMs, we set the first-order Taylor approximation of $\GradientLVk$ around the current estimate $\widetilde{\vB}_k$ to zero and derive the update for $\vB_k$ from that equation, i.e.\ via
\begin{align*}
\mathbf{0} = \GradientLVk &  \approx \GradientLHatVk +  \HessianLHatVk \; (\vB_k - \widetilde{\vB}_k)\\
\vB_k  &  \approx \widetilde{\vB}_k - \HessianInvLHatVk \; \GradientLHatVk,
\end{align*}
where
\begin{itemize}
	\item $\widetilde{\vB}_k $ is the current estimate of $\vB_k$;
	\item $\HessianLHatVk$ is the Hessian matrix containing all the second-order partial derivatives of $l(\etaB)$ w.r.t.\ $\vB_k$ evaluated at $\widetilde{\vB}_k $ (or some approximation thereof); and
	\item $\GradientLHatVk$ is the gradient vector containing the first-order partial derivatives of $l(\etaB)$ w.r.t.\ $\vB_k$ evaluated at $\widetilde{\vB}_k $.
\end{itemize} 
Since $\etaB$ is the weighted sum of transformed predictors, i.e.\ $\textstyle{\etaB =  \sum_{k=0}^p \beta_k  \GB_k \vB_k}$, the gradient vector of $\etaB$ w.r.t.\ $\vB_k$ is $\beta_k \GB_k$. Hence, from the chain rule
\begin{displaymath}
\GradientLHatVk = (\widetilde{\beta}_k \GB_k)^T \; \GradientLHatEta
\end{displaymath}
and
\begin{displaymath}
\HessianLHatVk = (\widetilde{\beta}_k \GB_k)^T \; \HessianLHatEta \; \widetilde{\beta}_k \GB_k,
\end{displaymath}
with $\textstyle{\GradientLHatEta }$ the gradient vector and $\textstyle{\HessianLHatEta}$  the Hessian matrix of $\textstyle{ l(\etaB) }$  w.r.t.\ $\etaB$ evaluated at current estimate $\widetilde{\etaB}$. Hence, the quantifications for all predictors with a nonnumeric scaling level are updated as
\begin{align}
\widetilde{\vB}_k^{+} &= \widetilde{\vB}_k - \HessianInvLHatVk \; \GradientLHatVk \nonumber \\
& = \widetilde{\vB}_k - \left\{(\widetilde{\beta}_k \GB_k)^T \; \HessianLHatEta \; \widetilde{\beta}_k \GB_k \right\}^{-1} \; (\widetilde{\beta}_k \GB_k)^T  \; \GradientLHatEta.
\label{GLM-OS_eq:VkUpdatesGLM-OS_InTermsOfEta}   
\end{align}
These updates are the unrestricted estimates of the quantifications and are the optimal solution for a  nominal scaling level. For the other scaling levels, restrictions have to be applied to $\widetilde{\vB}_k^{+}$ by fitting a nonmonotone or monotone step or spline function, as is done in OS-regression. Then the quantifications are standardized to ensure a unique solution. 

\vspace{\baselineskip}
After updating the quantifications $\vB_k$ for predictor $k$, $\beta_k$ needs to be updated accordingly. Again, updates can be derived via the first-order Taylor approximation of $\GradientLBetak$, which results in
\begin{equation}
\widetilde{\beta}_k^{+} = \widetilde{\beta}_k - \HessianInvLHatBetak \; \GradientLHatBetak,
\end{equation}
where 
\begin{itemize}
	\item $\widetilde{\beta}_k$ is the current estimate of $\beta_k$;
	\item $\HessianLHatBetak$ is the Hessian matrix containing all the second-order partial derivatives of $l(\etaB)$ w.r.t.\ $\beta_k$ evaluated at $\widetilde{\beta}_k$ (or some approximation thereof); and
	\item $\GradientLHatBetak$ is the gradient vector containing the first-order partial derivatives of $l(\etaB)$ w.r.t.\ $\beta_k$  evaluated at $\widetilde{\beta}_k$.
\end{itemize}
Using the chain rule,
\begin{align*}
\widetilde{\beta}_k^{+} & = \widetilde{\beta}_k - \HessianInvLHatBetak \; \GradientLHatBetak \\
& = \widetilde{\beta}_k - \left\{(\GB_k \widetilde{\vB}_k)^T \; \HessianLHatEta \; \GB_k \widetilde{\vB}_k \right\}^{-1} \; (\GB_k \widetilde{\vB}_k)^T  \; \GradientLHatEta, 
\end{align*}
where $\textstyle{\GradientLHatEta }$ and $\textstyle{\HessianLHatEta}$ are recalculated in between updating $\widetilde{\vB}_k$ and $\widetilde{\beta}_k$.

\vspace{\baselineskip}
The modified version of the Newton-Raphson method for GLM-OS can be summarized as follows.

\begin{framed}
	\noindent \textbf{GLM-OS algorithm:}
	\begin{description}
		\item[Initialization:] Set $\GB_0 = \boldsymbol{1}_n$ and $\vB_0 = \{1\}$, create $\GB_1, \ldots, \GB_p$ based on the data, and initialize the model parameters $\widetilde{\beta}_0, \ldots, \widetilde{\beta}_p$ and $\widetilde{\vB}_1, \ldots, \widetilde{\vB}_p$.
		
		\item[Cycle:] For $k=0, \ldots, p$, do:
		\begin{description}
			\item[Step 1:] Calculate the Hessian matrix $\textstyle{ \HessianLHatEta }$ and the gradient vector $\textstyle{ \GradientLHatEta }$.
			
			\item[Step 2:] If the scaling level of variable $k$ is nonnumeric, calculate the unrestricted estimates of the quantifications of $k$ as 
			\begin{displaymath}
			\widetilde{\vB}_k^{+} = \widetilde{\vB}_k - \left\{(\widetilde{\beta}_k \GB_k)^T \; \HessianLHatEta \; \widetilde{\beta}_k \GB_k \right\}^{-1} \; (\widetilde{\beta}_k \GB_k)^T  \; \GradientLHatEta.
			\end{displaymath}
			
			Apply appropriate scaling restrictions to $\widetilde{\vB}_k^{+}$ and standardize the result. 
			
			\item[Step 3:] Update the Hessian matrix $\textstyle{ \HessianLHatEta }$ and the gradient vector $\textstyle{ \GradientLHatEta }$ using the current estimate $\widetilde{\vB}_k$.
			
			\item[Step 4:] Update the estimate for model coefficient $\beta_k$ as 
			\begin{displaymath}
			\widetilde{\beta}_k^{+} = \widetilde{\beta}_k - \left\{(\GB_k \widetilde{\vB}_k)^T \; \HessianLEta \; \GB_k \widetilde{\vB}_k \right\}^{-1} \; (\GB_k \widetilde{\vB}_k)^T  \; \GradientLHatEta.
			\end{displaymath}
			
		\end{description}
		\item[Convergence:] Repeat the cycle until convergence criteria are met.

	\end{description}
\end{framed}

\subsection{The relation between the Newton-Raphson method for GLM(-OS)s, IRLS, and OS-regression}
The Newton-Raphson method for GLMs is often referred to as Iterative Reweighed Least Squares (IRLS), because the algorithm iteratively solves reweighted least squares problems. This will be explained in this section. This relation between GLM estimation and least squares is important for GLM-OS, because it accommodates the use of monotone regression and I-splines when finding optimal quantifications.

As was shown in \autoref{GLM-OS_subsec:NewtonRaphsonForOrdinaryGLM}, $\widetilde{\betaB}$ in an ordinary GLM is updated in each iteration as
\begin{displaymath}
\widetilde{\betaB}^{+} = \widetilde{\betaB} - \HessianInvLHatBeta \; \GradientLHatBeta.
\end{displaymath}
Given that the linear combination in ordinary GLMs is $\textstyle{\etaB =  \XB \betaB }$, the matrix containing all its partial derivatives w.r.t.\ $\beta_1, \ldots, \beta_k$ is $\XB$. Hence, according to the chain rule 
\begin{displaymath}
\textstyle{ \HessianLHatBeta = \XB^T \; \HessianLHatEta \; \XB}
\end{displaymath}and
\begin{displaymath}
\textstyle{\GradientLHatBeta= \XB^T \; \GradientLHatEta }.
\end{displaymath}
Hence, the updates for the model parameters $\widetilde{\betaB}$ can be rewritten as
\begin{align*}
\widetilde{\betaB}^{+} 
& = \widetilde{\betaB} - \HessianInvLHatBeta \; \GradientLHatBeta\\
& = \widetilde{\betaB} - \left\{\XB^T \; \HessianLHatEta \; \XB \right\}^{-1} \; \XB^T \;\GradientLHatEta\\
&= \left\{\XB^T \; \HessianLHatEta \; \XB\right\}^{-1} \; \XB^T \; \HessianLHatEta \left\{ \XB \widetilde{\betaB} - \HessianInvLHatEta \; \GradientLHatEta \right\}\\
& = \left\{\XB^T \; \HessianLHatEta \; \XB\right\}^{-1} \; \XB^T \; \HessianLHatEta \; \zB
\end{align*}
where $\textstyle{\zB = \XB \widetilde{\betaB} - \HessianInvLHatEta \; \GradientLHatEta}$. These updates are exactly the solution to the weighted least squares problem
\begin{displaymath}
\text{argmin}_{\betaB}\; \lnorm \zB - \XB \betaB \rnorm ^2_{ \HessianLHatEta},
\end{displaymath}
with $ \HessianLHatEta$ the (diagonal) matrix with weights for each observation. Hence, the Newton-Raphson algorithm iteratively optimizes a weighted least squares problem in which the weights are updated in each iteration. For this reason, it is often called the Iterative Reweighted Least Squares algorithm.

\vspace{\baselineskip}

The same reasoning can be used to show that the GLM-OS algorithm iteratively solves the weighted least squares problems 
\begin{align}
\label{GLM-OS_eq:LossFunctionsOfGLM-OS_quantifications}
\text{argmin}_{\vB_k}  &\lnorm \zB_k - \widetilde{\beta}_k \GB_k \vB_k \rnorm ^2_{ \HessianLHatEta}
\end{align}
and
\begin{align}
\label{GLM-OS_eq:LossFunctionsOfGLM-OS_coefficients}
\text{argmin}_{\beta_k}  &\lnorm \zB_k - \beta_k \GB_k \widetilde{\vB}_k \rnorm ^2_{ \HessianLHatEta},
\end{align}
with $\zB_k = \widetilde{\beta}_k \GB_k \widetilde{\vB}_k - \HessianInvLHatEta \GradientLHatEta$. 

Since the GLM-OS alternates between updating the model coefficients and the quantifications, it could be referred to as the Iterative Reweighted Alternating Least Squares (IRALS) algorithm. Note that the weights in $\HessianLHatEta$ are recalculated between calculating the updates of $\widetilde{\vB}_k^+$ and $\widetilde{\beta}_k^+$, and that the objects should be weighted accordingly when fitting the step or spline functions.

Although loss functions \eqref{GLM-OS_eq:LossFunctionsOfGLM-OS_quantifications} and \eqref{GLM-OS_eq:LossFunctionsOfGLM-OS_coefficients} look very similar to loss function \eqref{GLM-OS_eq:LossFunctionOS-regression_MatrixNotation_withUk} of the OS-regression algorithm, they are different. In the GLM-OS setting the objects are weighted according to the Hessian entries, while in OS-regression they receive equal weights. Furthermore, in GLM-OS the least squares problems change at each iteration and are subproblems that serve as intermediate steps to get closer to the maximum of the (log-)likelihood. In OS-regression, the minimization of the loss function is the actual optimization problem. 

\subsection{Example: logistic regression with optimal scaling transformations}

The GLM-OS algorithm as described previously can be applied to a variety of GLMs. In this paper, we focus on the logistic regression model, which is used when the outcome of interest is dichotomous. It models the probability $\pi_i$ that observation $i$ has response $y_i=1$, given observed predictor values $\xB_i$ via the log of the odds. The weighted sum of (transformed) predicted variables $\eta_i$ is the weighted sum of the log odds and is converted to probabilities, i.e.\
\begin{equation}
\label{GLM-OS_eq:LinkFunctionLogReg}
\pi_i = \frac{1}{1 + \exp(-\eta_i)},
\end{equation}
which represents the probability of success ($y_i = 1$) in a Bernoulli trial. The resulting likelihood function is
\begin{equation}
L(\etaB) = \prod_{i=1}^n \pi_i^{y_i} (1-\pi_i)^{1-y_i}  = \prod_{i=1}^n \exp(\eta_i)^{y_i}   \frac{1}{1 + \exp(\eta_i)},
\end{equation}
with corresponding log-likelihood
\begin{equation}
\label{eq:LogisticRegression_logLikelihood}
l(\etaB) = \sum_{i=1}^n y_i \eta_i -  \sum_{i=1}^n \log\{1 + \exp(\eta_i )\}.
\end{equation}

We use the modified Newton-Raphson method as described in \autoref{GLM-OS_subsec:ModifiedNewtonRaphsonScoringMethod} to maximize \eqref{eq:LogisticRegression_logLikelihood} to find the optimal estimates for both the model parameters $\betaB$ and quantifications $\vB_1, \ldots, \vB_p$. To simplify later calculations, we recast the maximization problem into a minimization problem and find the minimum of the negative of the log-likelihood. 

To apply the algorithm, we need to derive the gradient vector $\textstyle{ \GradientNegLHatEta }$ and the Hessian matrix $\textstyle{ \HessianNegLHatEta }$ of $-l(\etaB)$ w.r.t.\ $\etaB$ evaluated at the current estimate $\widetilde{\etaB}$. The gradient is
\begin{displaymath}
\textstyle{ \GradientNegLHatEta} = \boldsymbol{\pi} - \yB
\end{displaymath} 
and the Hessian is
\begin{align*}
\HessianNegLHatEta = \text{diag} \left\{\boldsymbol{\pi} (1 - \boldsymbol{\pi} ) \right\},
\end{align*}
where $\boldsymbol{\pi}$ is the $n$-vector of probabilities $\pi_i$ as defined in \eqref{GLM-OS_eq:LinkFunctionLogReg}. Calculation details are provided in \autoref{OSforGLM_app:LikelihoodLogReg_withDerivatives}.

The updates for quantifications $\vB_k$ and coefficient $\beta_k$ in each iteration are as follows
\begin{equation}
\widetilde{\vB}_k^{+} = \widetilde{\vB}_k - \left[(\widetilde{\beta}_k \GB_k)^T \; \text{diag} \left\{\boldsymbol{\pi} (1 - \boldsymbol{\pi} ) \right\} \; \widetilde{\beta}_k \GB_k \right]^{-1} \; (\widetilde{\beta}_k \GB_k)^T  \; (\boldsymbol{\pi} - \yB)
\end{equation}
and
\begin{equation}
\widetilde{\beta}_k^{+} = \widetilde{\beta}_k - \left[(\GB_k \widetilde{\vB}_k)^T \; \text{diag} \left\{\boldsymbol{\pi} (1 - \boldsymbol{\pi} ) \right\} \; \GB_k \widetilde{\vB}_k \right]^{-1} \; (\GB_k \widetilde{\vB}_k)^T  \; (\boldsymbol{\pi} - \yB),
\end{equation}
where $\boldsymbol{\pi}$ is recalculated before updating $\widetilde{\beta}_k$.

This algorithm that integrates OS transformations in the logistic regression model has been implemented in R software environment ({\citet{Software:R}) to perform the analyses that will be described in the next section. This implementation of the algorithm speeds up the calculations and saves memory space by avoiding matrix multiplications with and storage of the sparse but potentially big matrices $\GB_k$.

	
	\section{Application of GLM with optimal scaling: logistic regression}
	
	In this section the GLM-OS method is applied to three different datasets. In all examples, we use a logistic regression model to predict a binary classification from a set of predictors. More specifically, we study how each predictor influences the odds of being classified in one of the classes given the other predictors, which we usually refer to as a change in probability or likelihood to be in that class. 	
	Each of the three illustration focuses on a particular predictor type, namely categorical, ordinal and mixed data, and on different scaling levels which can be used to analyze these types of data.
	
	\subsection{Transformation and visualization of categorical predictors}	\label{GLM-OS_subsec:NominalScalingVSDummyCoding} 
	We use a medical dataset to show how the OS methodology deals with categorical data by finding optimal quantifications for each category, using the nominal scaling level (i.e.\ nonmonotone step-functions). This approach is an alternative to the use of dummy variables, which is the standard approach for categorical predictors in GLMs. We will show how the replacement of dummy variables by nominal quantifications will simplify the visualization and interpretation of the model, while it also benefits the computational process. 
	
	\subsubsection{Data description}
	
	The first dataset is provided by the German multi-center project DINSTAP (Differentielle INdikationsstellung Station\"arer und TAgesklinischer Psychotherapie; differential indication for inpatient and day clinic psychotherapy). The aim of the original project was to explore which criteria are used by clinicians to choose between an inpatient or a day clinic psychotherapy treatment.
	
	Data on 25 possible predictors for treatment choice were collected. In the analysis illustrated in this section, we will only include the six most important variables for prediction (\citet{Art:DINSTAPdata_AnalyseAnitaEtAl}); namely \emph{Need for medical care}, \emph{Travel time}, \emph{Need for relief from family conflicts}, \emph{Need for relief from strain}, \emph{Psychological restrictions of mobility}, and \emph{Need to apply therapy in everyday life}. 
	
	Since this data analysis is for illustration purposes only, we focus only on the complete cases (n = 342). For 53.8\% of these patients, clinicians preferred a day clinic treatment ($y = 0$), while for the others (46.2\%) an inpatient treatment ($y = 1$) seemed more suitable. 
	
	We refer to \citet{Art:DINSTAPdata_originalArticle} for a description of the full dataset.

	\subsubsection{OS transformations with nominal scaling level}

	In the OS methodology, optimal quantifications for the categories of the predictor variables are found within the restrictions of the chosen scaling level. The least restrictive scaling level is a  nominal transformation in which no ordering of the categories is taken into account.  This scaling level is equivalent to the standard approach to handle categorical data, in which first dummy variables are defined to represent the categories and then model coefficients are estimated for each dummy individually. Namely, if there are $C_k$ categories for variable $k$, then $C_k-1$ dummies are defined and hence $C_k-1$ regression coefficients will be estimated, each indicating the effect of one category in comparison to left out (reference) category.

	In contrast, optimal scaling assigns quantifications to all categories and estimates a single regression coefficient for each categorical predictor. Namely, the vector $\vB_k$ of length $C_k$ contains quantifications for the $C_k$ categories and matrix $\GB_k$ contains $C_k$ columns representing all the categories, such that $\GB_k \vB_k$ gives the transformed predictor which is weighted by one regression coefficient $\beta_k$. If no restrictions (nominal scaling level) are applied to the quantifications $\vB_k$ it will give similar results as analysis on dummy variables, but these results are represented differently, as shown in \autoref{GLM-OS_tab:DummyCodingVSOSScaling}
	
	\begin{table}[h]
		\centering
		\begin{tabular}{ccc}
			\toprule
			Category & Dummy coding & Optimal scaling\\
			\midrule
			1 &  & $\beta_k v_{k_1}$ \\
			2 & $\beta_{k_2} $& $\beta_k v_{k_2}$ \\
			\vdots & \vdots & \vdots \\
			$C_k -1$ & $\beta_{k_{C_k -1}} $& $\beta_k v_{k_{C_k -1}}$ \\
			$C_k$ & $\beta_{k_{C_k}} $& $\beta_k v_{k_{C_k}}$ \\
			\bottomrule
		\end{tabular}
		\caption{Contributions of the $C_k$ levels of a categorical variable $k$ to the linear combination of predictor variables for the ordinary regression model with dummy coding and the optimal scaling model.}
		\label{GLM-OS_tab:DummyCodingVSOSScaling}
	\end{table}

	The OS transformations for the six predictor variables are visualized by plotting the estimated quantifications against the original values of the categories (\autoref{GLM-OS_fig:DINSTAPanalsys_quantifications}) and the estimated regression coefficients are given below each plot. 
	
	The lines that connect the dots have no meaning since there are no intermediate categories. However, their slopes visualize useful additional information about the relation between levels. For example, a steep slope indicates a large difference between consecutive categories, while no or a small increase or decrease is indicated by a flat or near to flat upward or downward slope. In this way, the lines help interpreting the result and are therefore included in the plots. 
	
	\begin{figure}[h]
		\centering
		\includegraphics[width=\linewidth]{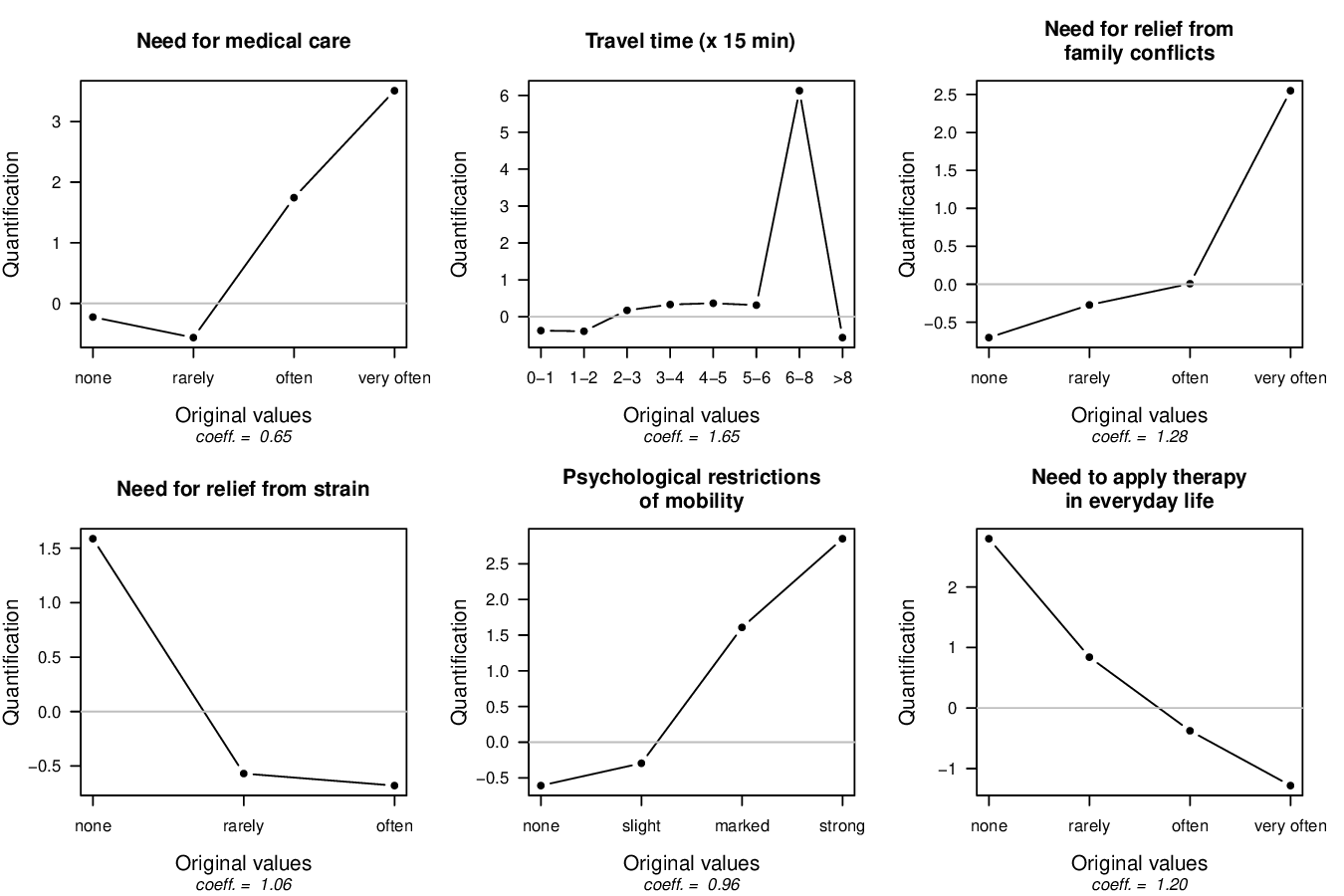}
		\caption{DINSTAP data: Nominal quantifications estimated for each of the original categories of the six predictor variables. Regression coefficients are provided below the plots of the corresponding predictor. The estimated intercept is 0.56.}
		\label{GLM-OS_fig:DINSTAPanalsys_quantifications}
	\end{figure}
	
	\vspace{\baselineskip}
	The interpretation of the influence of a specific variable $k$ in a model with OS transformations is via the estimated model coefficient $\widehat{\beta}_k$ which by its magnitude indicates the strength of the effect, and via the estimated quantifications $\widehat{\vB}_k$ which indicate the shape of the effect. The sign of $\widehat{\beta}_k$ combined with the signs of $\widehat{\vB}_k$ show the direction(s) between categories.
	
	To understand which predictors have the strongest effect on the outcome, we first compare the regression coefficients. Given the values of the estimated coefficients, the predictors can be ordered according to the strength of their effect; i.e.\ \emph{Travel Time} has the strongest effect ($\widehat{\beta} = 1.65$), followed by \emph{Need for relief from family conflicts} ($\widehat{\beta}  = 1.28$), \emph{Need to apply therapy in everyday life} ($\widehat{\beta}  = 1.20$), \emph{Need for relief from strain} ($\widehat{\beta}  = 1.06$), \emph{Psychological restrictions of mobility} ($\widehat{\beta}  = 0.96$), and \emph{Need for medical care} ($\widehat{\beta}  = 0.65$). 
	The relative proportions of the model coefficients can also be used to draw conclusions. For example, we can conclude that the effect of \emph{Need for relief from family conflicts} is twice as big as the effect of \emph{Need for relief for medical care} ($\widehat{\beta}  = 1.28$ vs. $\widehat{\beta}  = 0.65$).

	The direction(s) of the effect between categories is given by the combination of its quantification and the sign of the predictor's model coefficient. 
	For example, the large positive quantifications of the third category (often) of \emph{Need for medical care} in combination with the positive model coefficient of this predictor indicates that if medical care is often required, a patient is more likely to be referred to inpatient treatment than day clinic treatment (inpatient is coded higher than day clinic). Furthermore, this probability increases if medical care is very often needed (fourth category). However, when no medical care is required (first category) or just rarely (second category), this will hardly influence a clinician's choice. A similar pattern is seen for \emph{Psychological restrictions of mobility}.
	Additionally, only a strong need to be relieved from family conflicts seems to be a reason to choose for an inpatient treatment.
	Apparently inpatient treatment is believed to give additional mental stress, because this type of treatment is usually only given when there is no need for relief from strain. 
	Moreover, the effect of the need to apply the therapy in everyday life seems to be almost linear in its categories. 
	Surprisingly, a \textit{Travel Time} between 6 and 8 quarters of an hour seems to be a strong indicator for inpatient treatment, while an even longer travel time is an indicator for day clinic treatment. This is a questionable result which is due to the small number of patients in these two categories (8 and 6 patients respectively), and inpatient treatment being chosen for all eight patients in this category.

	\vspace{\baselineskip}
	Concluding, the visualizations of the quantifications help to interpret the results. A closer look at their exact values and the model coefficients will give a more detailed interpretation.

	\FloatBarrier 
	\subsubsection{Comparison between OS transformations and the use of dummy variables}
	In standard logistic regression dummy variables are created to estimate the effects of each category for all predictors on the outcome. The analysis on the DINSTAP data gives the estimates in \autoref{GLM-OS_tab:CoefficientsNormalLogReg_DINSTAP}.

	These estimates should always be interpreted in terms of the reference category, thus the estimate $-0.220$ for category 2 of \emph{Need for medical care} indicates that for patients who are classified in the second category of this predictor, the weighted sum of predictors is 0.220 lower than the weighted sum for those in the first category (= reference category). 
	To compare the second and third categories, it is necessary to subtract the corresponding coefficients. Hence, to know whether being classified in category \emph{rarely} instead of \emph{often} or in \emph{often} instead of \emph{very often} has a bigger effect on the treatment choice, we have to compare the differences between their corresponding coefficients. Since $1.284 - (-0.220) = 1.504$ and $2.434 - 1.284 = 1.150$, this implies that the step from the second to the third category is larger than the step from the third to the fourth level. The same conclusion could be drawn by looking at the slopes in the quantification plots in \autoref{GLM-OS_fig:DINSTAPanalsys_quantifications}.

	\begin{table}[h]
		\centering
		\begin{tabular}{p{4cm}rrrrrrrr}
			\toprule
			Category & 1/ref & 2 & 3 & 4 & 5 & 6 & 7 & 8 \\
			\midrule
			\makecell[l]{Intercept} & 3.34 &  &  &  &  &  &  &  \\[1em]
			\makecell[l]{Need For\\Medical Care} &  & -0.22 & 1.28 & 2.43 &  &  &  &  \\[1em]
			\makecell[l]{Travel Time} &  & -0.03 & 0.91 & 1.17 & 1.22 & 1.14 & 17.10 & -0.31 \\[1em]
			\makecell[l]{Need For Relief\\From Family Conflicts} &  & 0.55 & 0.91 & 4.17 &  &  &  &  \\[1em]
			\makecell[l]{Need For Relief\\From Strain} &  & -2.29 & -2.40 &  &  &  &  &  \\[1em]
			\makecell[l]{Psychological Restrictions\\Of Mobility} &  & 0.30 & 2.13 & 3.32 &  &  &  &  \\[1em]
			\makecell[l]{Need To Apply Therapy\\In Every Day Life} &  & -2.34 & -3.80 & -4.88 &  &  &  &  \\[1em]
			\bottomrule
		\end{tabular}
		\caption{DINSTAP data:
			Estimated model coefficients of the logistic regression model with dummy variables representing all categories of the categorical predictors.}
		\label{GLM-OS_tab:CoefficientsNormalLogReg_DINSTAP}
	\end{table}

	\vspace{\baselineskip}
	The equivalence between the results obtained with optimal scaling and the use of dummy variables can be seen through the differences between the category quantifications. For example, the difference in the effect of categories 1 and 2 of \emph{Need for Medical Care} in the optimal scaling result is the difference between the quantifications multiplied by the coefficient, $ 0.65 \cdot \{-0.563 - (-0.225 )\} = -0.220$, which is precisely the coefficient for the corresponding dummy variable.
	
	Hence, the results from ordinary logistic regression with dummy variables are essentially equal to the results of logistic regression with nominal scaling transformations, but they are represented differently. While the result for dummy variables focuses on the numeric coefficients only, OS puts more emphasis on visualization to improve the understanding of the quantification result, and provides regression coefficients for the predictors. The coefficients estimated for each dummy variable could also be plotted and the resulting figures will be very similar to those in \autoref{GLM-OS_fig:DINSTAPanalsys_quantifications}. However, most statistical software do not provide these plots as a default.

	\FloatBarrier 
	\subsection{Monotone transformations to facilitate interpretation}
	\label{GLM-OS_sec:MonotoneTransformationsForInterpretation}

	In the next illustration we use survey data to show the differences between nonmonotone and monotone quantifications for both ordered categorical and continuous data. If the prediction accuracy is not reduced significantly, it may be beneficial to put monotonicity constraints on the transformations.

	\subsubsection{Data description}

	For this illustration we use a subset of the 1987 National Indonesia Contraceptive Prevalence Survey data (\citet{Art:ContraceptiveMethodChoice_ArticleReference}, available from the ICU Machine Learning Repository via \url{https://archive.ics.uci.edu/ml/datasets/Contraceptive+Method+Choice}). The dataset contains several variables collected from married couples and their choice of contraceptive method. The categories of the outcome variable are no, long-term, or short-term use, and we merged the short- and long-term use into one category to create a binary outcome variable indicating whether couples use contraceptive methods ($y = 1$) or not ($y=0$). There are nine predictor variables of which three are binary, four are categorical with ordered levels, and two are continuous. There are no missing values for any of the variables ($n = 1472$).

	\subsubsection{Nonmonotone vs. monotone quantifications}
	
	Since the values for most predictor variables in this dataset are ordered (namely for four categorical and two continuous variables), this dataset is suitable to compare nonmonotone and monotone transformations. For the categorical variables, either a nonmonotone or monotone step function are fitted, and for the continuous variables we use a (non)monotone spline transformation (of degree two with one interior knot). The results are shown in \autoref{GLM-OS_fig:MonotoneAndNonMonotoneQuantifications_CMCdata}.

	Most estimated transformations are monotone even without imposing monotonicity, therefore the quantifications of the monotone and nonmonotone analyses are very similar for most predictors. The largest differences can be found for the variables \emph{Education Husband}, \emph{Occupation Husband}, and \emph{Number of previous children}. However, although the results are very similar, it may still be beneficial to apply the monotonicity constraints, since it may simplify the interpretation of the result or reduce overfitting (better EPE).
	
	For example, if there are no monotonicity restrictions, the model indicates that, given that all other variables are constant, if the husband is in the highest category of education, then the couple is less likely to use contraceptive methods, compared to the two middle categories. If monotonic restrictions are imposed, the quantifications of the three highest categories are equal and very close to zero. From this example we see that monotonicity may simplify the interpretation of the quantifications, since there is no need to explain the decrease at the end. A similar reasoning can be used for the quantifications of the \emph{Number of previous children}. Namely, with a monotone restriction, the slight dip around 9-11 children is smoothed out, which simplifies interpretation without loosing model fit. Hence, although differences with the nonmonotonic results are small, the monotonic quantifications of \emph{Education Husband}, \emph{Occupation Husband}, and \emph{Number of previous children} are easier to interpret and correspond more to reality than the nonmonotonic ones.

	\begin{figure}[h]
		\centering
		\includegraphics[width=\linewidth]{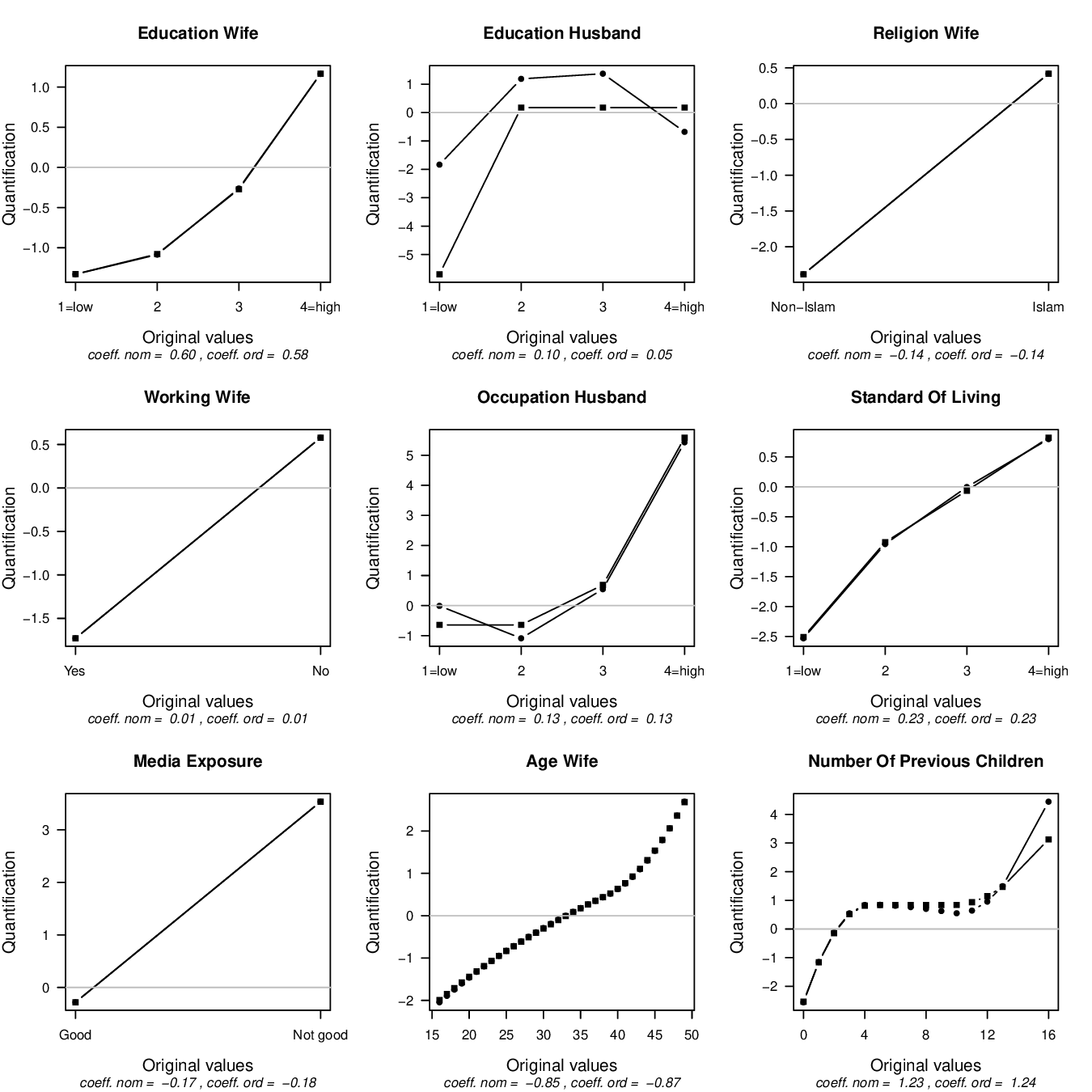}
		\caption{Contraceptive method choice data: Nominal (circles) and ordinal (squares) quantifications estimated for each of the original categories of the nine predictor variables. Categorical variables are transformed using step functions and continues predictors are transformed using splines. Estimated regression coefficients are provided below the plots of the corresponding predictor. The estimated intercepts are 0.30 for both the nonmonotone and monotone analyses.
		}
		\label{GLM-OS_fig:MonotoneAndNonMonotoneQuantifications_CMCdata}
	\end{figure}

		Also, in the case of low category frequencies, especially with probability 1, a more restrictive transformation can be beneficial for the EPE as it prevents overfitting. 
	
	\vspace{\baselineskip}
	Even though monotonic constraints ease interpretation, imposing too many restrictions on the transformations may hide the true relation between the predictor and outcome variables. Therefore it is important to check the model's performance for future observations before choosing for monotone scaling levels. This check can be done with cross-validation (CV). The results for this dataset are shown in \autoref{GLM-OS_tab:CVresults_CMC}.
	
	As can be expected, the prediction errors based on the test data (EPE) using a 10-fold cross-validation are higher for both models compared to the apparent prediction error (APE) calculated on the training data. The increase is slightly smaller for the model with monotone transformations, but the difference between the models is very small (0.187 vs.\ 0.186). This suggests that applying monotonicity does not hide any important relation between the predictors and the outcome variable.
	
	\begin{table}[h]
		\centering
		\begin{tabular}{rrrrr}
			\toprule
			& APE & EPE & SE(EPE) & MCR (\%)\\ 
			\midrule
			{Logistic Regression (linear)} & 0.205 & 0.211 & 0.0047 & 33.1 \\ 
			{GAM (nonmonotone)} & 0.181& 0.187 & 0.0052 & 28.1 \\ 
			GLM-OS (nonmonotone) & 0.181 & 0.187 & 0.0053 & 27.9 \\ 
			GLM-OS (monotone) & 0.181 & 0.186 & 0.0053 & 27.9 \\ 
			\bottomrule
		\end{tabular}
		\caption{Contraceptive method choice data: Apparent prediction error (APE) for the GLM-OS model with nonmonotone and monotone transformations, together with the 10-fold cross validation results: Expected Prediction Error (EPE) along with its standard error (SE(EPE)) and the Misclassification Rate (MCR). Results from standard logistic regression and GAM are added for comparison.}
		\label{GLM-OS_tab:CVresults_CMC}
	\end{table}
	
	\subsubsection{Relation with ordinary logistic regression and GAMs}

	In ordinary logistic regression, categorical predictors are included in the model by defining the categories with dummy variables and by analyzing continuous data linearly. For categorical data, the dummy coding essentially gives the same result as transformations with a nominal scaling levels, although the result is represented differently (see \autoref{GLM-OS_subsec:NominalScalingVSDummyCoding}). The main difference between the ordinary logistic regression and nonmonotone GLM-OS results for this dataset is in the restrictions applied on the continuous variables. Namely, the linear relations assumption for the logistic regression is more restrictive than the nonmonotone spline transformations in the GLM-OS. 
	
	In a GAM analysis, categorical variables are represented as dummy variables, as in ordinary logistic regression. Continuous variables are usually transformed using a nonmonotone spline function, but the algorithm to find the optimal spline is different from the algorithm used in OS. 
	Therefore, the objective of GAMs is similar to nonmonotone GLM-OS, but the results for categorical data are represented differently and the nonmonotone splines are fitted in a slightly different way. 
	
	In \autoref{GLM-OS_tab:ClassificationModelsAccordingToRestrictions} restrictions for ordinary logistic regression, (non)monotone GLM-OS, and GAMs are provided for comparison of the models.
	
			\begin{table}
		\centering
		\small
		\begin{tabular}{p{2cm}p{6cm}p{6cm}}
			\toprule
			Restrictions & Categorical predictors & Numeric predictors \\
			&  &  (or with many categories) \\
			\midrule 
			none & \textbullet~Logistic regression with dummies & \\
			& \textbullet~GAMs with dummies  &  \textbullet~GAMs with nonmonotonic splines \\
			&  \textbullet~GLM-OS with nonmonotonic step functions & \textbullet~GLM-OS with nonmonotonic splines \\[1em]
			monotonic & \textbullet~GLM-OS with monotonic step functions & \textbullet~GLM-OS with monotonic splines\\[2em]
			linear &  &\textbullet~Logistic regression\\
			&& \textbullet~GAMs with linear transformations \\
			&& \textbullet~GLM-OS with linear transformations\\
			\bottomrule
		\end{tabular}
		\caption{Classification of the four models according to the restrictions applied to categorical and numeric data.\\
		}	\label{GLM-OS_tab:ClassificationModelsAccordingToRestrictions}
	\end{table}

	\vspace{\baselineskip}
	The similarity of GAM and nonmonotonic GLM-OS is confirmed by the cross-validation results provided in \autoref{GLM-OS_tab:CVresults_CMC}. The small difference between the fitted splines have little influence on the predicted values of the observations. 
	
	Larger differences are seen for ordinary logistic regression. The prediction errors and misclassification percentages for the classic analysis are higher than those for (non)monotone GLM-OS and GAM. This suggests that the linear relations assumption seems too strict for the continuous variables in this dataset. Hence, imposing monotonicity will enhance interpretation, but imposing linear relations (which would simplify the interpretation even more) will hide nonlinear relations between the continuous predictors and outcome that are important for prediction.

	\FloatBarrier 
	\subsection{Mixed scaling levels}
	\label{GLM-OS_subsec:MixedScalingLevels} 
	Although monotone quantifications are usually easier to interpret, choosing a monotone scaling level is only correct if the predictor is measured on at least an ordinal scale. For example, it would not make sense to impose monotonicity on the transformations of nominal categorical variables like countries, color, or blood type. On the other hand, imposing no restrictions can be suitable for an ordered variable (categorical or continuous) if a nonmonotone relation is expected between this variable and the outcome, or if one has no idea about the relation and wants this to be revealed by the analysis. Therefore, it is important to choose each scaling level either in accordance with the measurement level of the predictor if its relation is known or assumed a priori, or choose the least restrictive level to explore its relation.
	
	In OS a different scaling level can be selected for each individual predictor. Usually this results in an analysis with a mix of scaling levels most suitable for the data. We will illustrate a mixed scaling level model with a medical dataset. 
	
	\subsubsection{Data description}

	\begin{figure}[h]
		\centering
		\includegraphics[width=\linewidth]{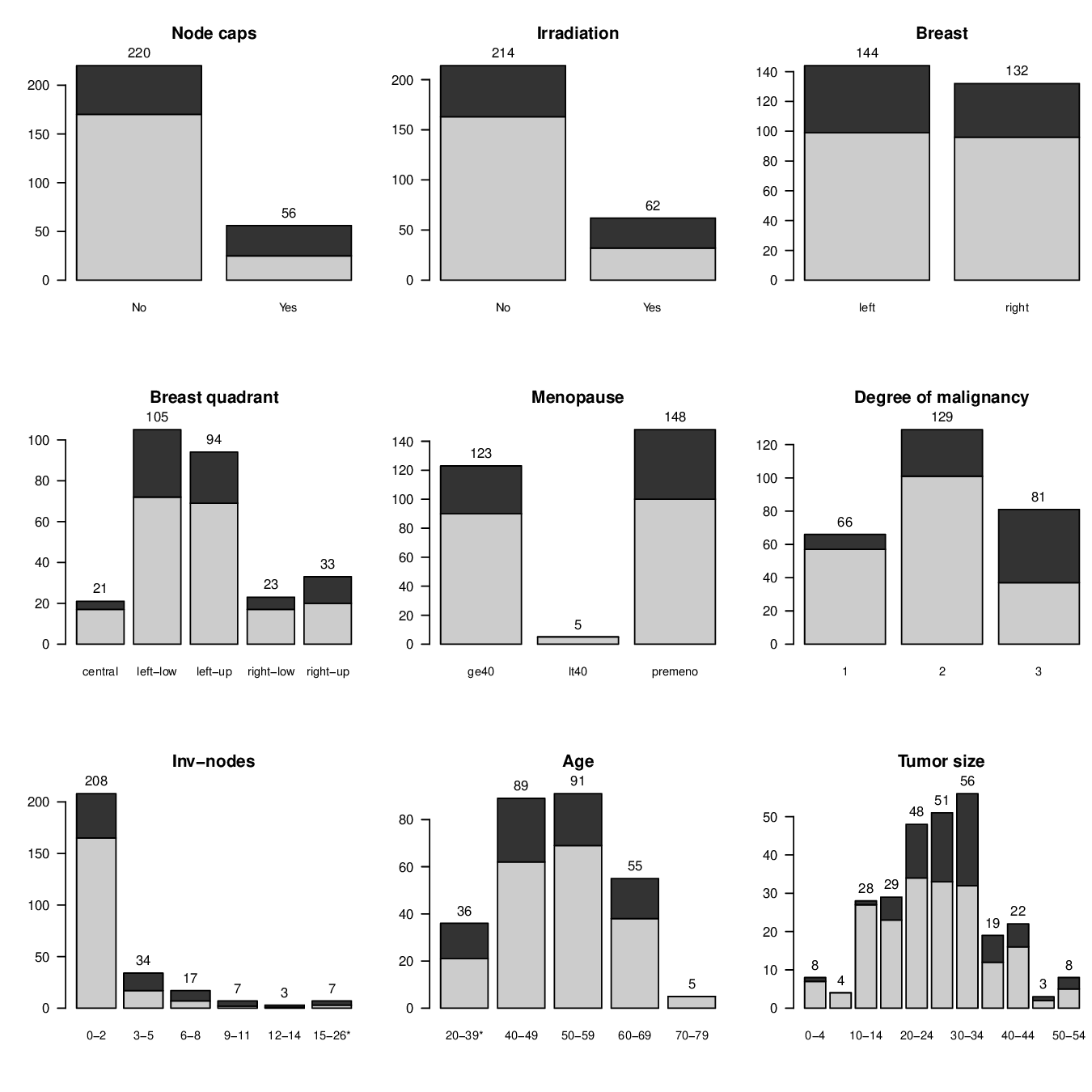}
		\caption{Breast cancer recurrence data: Distribution of observation over the categories of the predictor variables, split by outcome value (dark grey = recurrence events, light grey = no recurrence events).\\
			*This category is a union of two categories that were merged because one of the original categories contained only one observation. }
		\label{GLM-OS_fig:DistributionOfObservationsOverCategories_BreastCancerRecurrenceData}
	\end{figure} 
	
	For this illustration we use the breast cancer recurrence dataset (M.\, Zwitter \& M.\, Soklic, University Medical Center, Institute of Oncology, Ljubljana, Yugoslavia; available from the ICU Machine Learning Repository via \url{https://archive.ics.uci.edu/ml/datasets/breast+cancer}). This dataset contains information on the binary response variable which indicates whether a patient experienced recurrence-events ($y=1$) or not ($y = 0$). The aim is to predict the probability of recurrence-events from nine categorical and continuous predictor variables. 
	
	The predictor variables were measured on different scales. Variables \emph{Node caps}, \emph{Irradiation}, and \emph{Breast} are categorical with two unordered categories. \emph{Breast quadrant} and \emph{Menopause} are categorical with more than two unordered categories. The \emph{Degree of malignancy} is indicated by three levels. These levels are ordered and a higher level indicates more abnormal cells. Finally, there are three continuous predictors that were discretized into categories; \emph{Inv-nodes}, \emph{Age}, and \emph{Tumor size}. Unfortunately, the dataset does not include the original continuous values, so we can only use the discretized results.
	
	The dataset contains 276 complete cases and the distribution of these observations over the predictors' categories is shown in \autoref{GLM-OS_fig:DistributionOfObservationsOverCategories_BreastCancerRecurrenceData}.

	\subsubsection{GLM-OS with scaling levels according to measurement level}
	Given that all predictors have different measurement levels, they require a different type of transformation in the logistic regression analysis. In the OS setting this can easily be done by selecting an appropriate scaling level for each variable. 
	
	For binary variables, all scaling levels will result in the same quantifications, namely the ones resulting from numeric scaling level (because it only has two values, so there is always one increasing or decreasing linear line in the transformation plot between these values, no matter the scaling level). Hence, for the three binary predictors, any scaling level could be chosen. To reduce calculation time it is best to choose the numeric scaling level. 
	
	The categorical variables \emph{Breast quadrant}, \emph{Menopause}, and \emph{Degree of malignancy} contain up to five categories. Since the levels of the first two predictors are unordered, a nonmonotone step function is the most appropriate. For \emph{Degree of malignancy} a monotone step function is more suitable since its levels are ordered and we want to retain this ordering in the quantifications.
	
	The last three continuous predictors were binned into categories; 6 and 5 categories respectively for predictors \emph{Inv-nodes} and \emph{Age} and 11 for\emph{Tumor size}. Restrictions for a monotone step function are used to fit the quantifications for \emph{Inv-nodes} and \emph{Age} and we choose a smooth transformation by fitting a monotone spline for \emph{Tumor size} (quadratic, 1 interior knot).

	\begin{figure}[h]
		\centering
		\includegraphics[width=\linewidth]{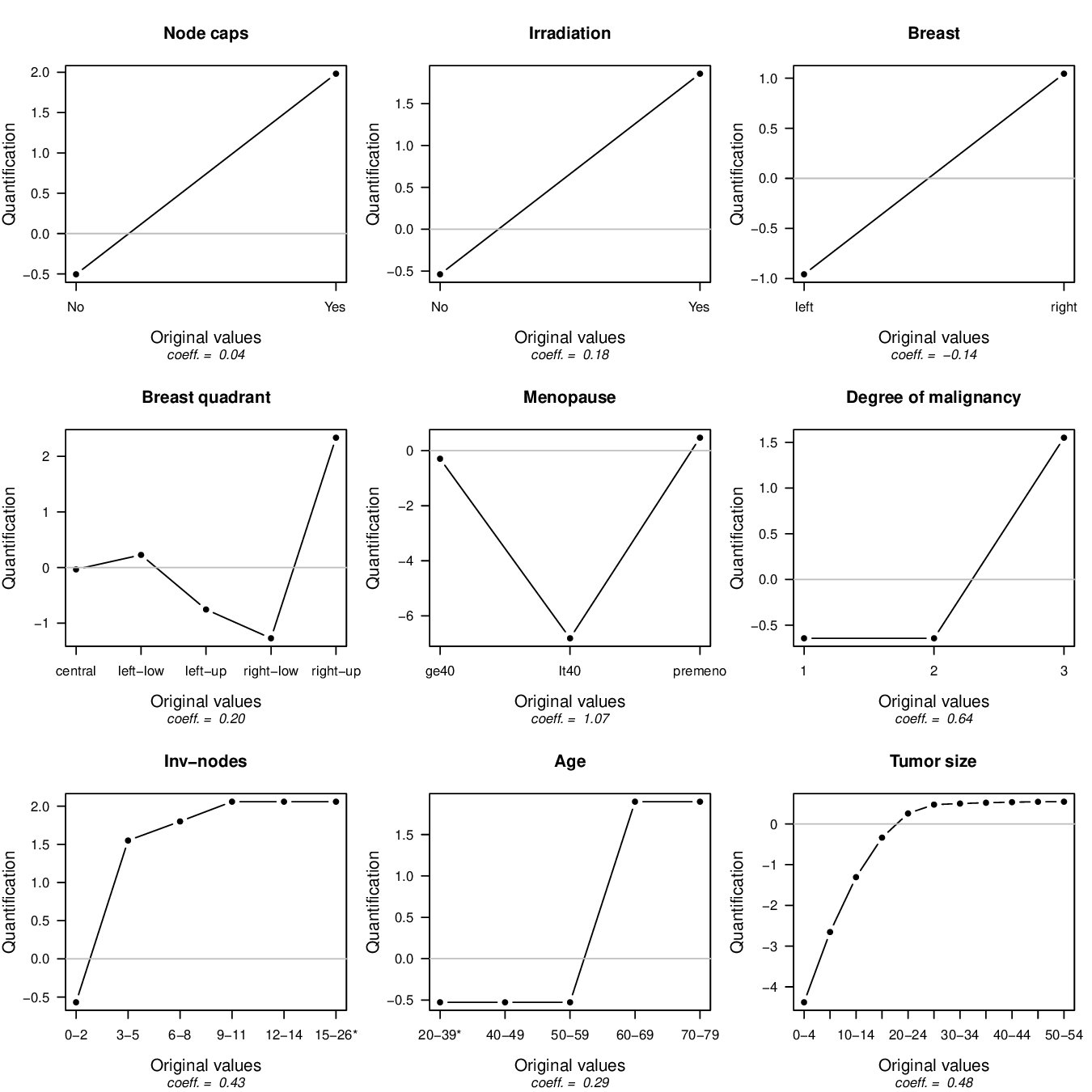}
		\caption{Breast cancer recurrence data: Quantifications estimated for each of the original categories of the nine predictor variables. Unordered categorical variables are transformed using nonmonotone step functions and ordered categorical variables are transformed using monotone step functions. Variable \emph{Tumor size} is transformed using a monotone spline. Regression coefficients are provided below the plots of the corresponding predictor. The estimated intercept is -1.24. \\
			*This category is a union of two categories that were merged because one of the original categories contained only one observation. 
		}
		\label{GLM-OS_fig:MixedLevelQuantifications_BreastCancerData}
	\end{figure}
	
	\vspace{\baselineskip}
	Results based on the logistic regression analysis with OS transformations are given in 	\autoref{GLM-OS_fig:MixedLevelQuantifications_BreastCancerData}. Several conclusions can be drawn from this exploratory analysis. 
	
	The values of the estimated coefficients suggest that whether the cancer metastasizes to a lymph node (\emph{Node caps}) has little influence on the probability of a recurrence-event; nor does the use of irradiation therapy, nor the tumor location (indicated by \emph{Breast} and \emph{Breast Quadrant}). A small effect is seen for \emph{Degree of malignancy}, \emph{Inv-nodes}, \emph{Age} and \emph{Tumor size}. The highest effect is for \emph{Menopause}, with its quantifications indicating that the \emph{lt40} stage is protective against recurrence-events. However, the quantified value of this stage of about -6 is outlying (remember the quantifications are $Z$-scores) due to a low frequency (5), all with the same outcome (see corresponding bar plot in \autoref{GLM-OS_fig:DistributionOfObservationsOverCategories_BreastCancerRecurrenceData}), so more information should be collected from patients in this menopause stage to verify this result.
	
	The ordinal predictors seem quite informative. For example, patients who were in the third degree of malignancy were more likely to get recurrence-events compared to those who were in one of the two lower levels. 
	Furthermore, the transformations of the binned continuous predictors indicate that recurrence-events are more likely to occur if lymph nodes contain metastatic breast cancer ($>2$ \emph{Inv-nodes}), or if the tumor size was above 20. Note that although the probability of a recurrence-event increases with tumor size 0 to 20, it barely increases thereafter, so up to a size of 20 the effect is linear and levels off after that size. Moreover, especially women of age 60 or older are more likely to experience recurrence-events.

	\subsubsection{Comparison with nonmonotone scaling level and linearity restrictions}

	In the current analysis, all scaling levels are chosen to preserve all properties of the data (i.e.\ only grouping for non-ordered categorical variables, plus ordering for ordered categorical variables; since the continuous variables were binned, they are to be regarded as ordered categorical). However, scaling levels with less restrictions may be chosen. For example, although \emph{Age} was transformed according to an ordinal scaling level, because we wanted to preserve the ordering, it might be that Age is not monotonically  related to the probability of recurrence-events. Therefore, we may check whether a nonmonotone scaling level is more suitable for an ordinal predictor as well. 
	
	A cross-validation is used to compare the prediction accuracies of the previous and less restrictive models with only nonmonotone transformations. In the latter model, nonmonotone step functions were used to transform all variables except for \emph{Tumor size}, for which a nonmonotone spline function (quadratic, 1 interior knot) was chosen. Results are shown in \autoref{GLM-OS_tab:CVresults_BreastCancer}.
	
	\begin{table}[h]
		\centering
		\begin{tabular}{p{7cm}p{0.8cm}p{0.8cm}p{1.2cm}p{1.2cm}}
			\toprule
			& APE & EPE & SE(EPE) & MCR(\%) \\ 
			\midrule		
			\raggedleft GLM-OS (nonmonotone) & 0.156 & 0.195 & 0.0157 & 27.5 \\ 
			\raggedleft GAM (nonmonotone) & 0.150 & 0.193 & 0.0157 & 27.5 \\ 
			\raggedleft Logistic regression (1 variable linear) & 0.154 & 0.191 & 0.0156 & 26.1 \\ 
			\raggedleft Logistic regression (4 variables linear) & 0.166 & 0.188 & 0.0138 & 28.3 \\ 
			\raggedleft GLM-OS (mixed scaling levels) & 0.156 & 0.180 & 0.0142 & 26.4 \\ 
			\bottomrule
		\end{tabular}
		\caption{Breast cancer recurrence data: Apparent prediction error (APE) for the GLM-OS model with nonmonotone and monotone transformations, together with the 10-fold cross validation: Expected Prediction Error (EPE) along with its standard error (SE(EPE)) and the Misclassification Rate (MCR). Results from standard logistic regression and GAM are added for comparison.}
		\label{GLM-OS_tab:CVresults_BreastCancer}
	\end{table}
	
	Cross-validation shows that the analysis with monotonicity restrictions produce smaller expected prediction errors (EPE) and misclassification rate, while the apparent prediction error (APE) is the same. This result suggests that a nonmonotone approach yields overfitting. So, in addition to easing interpretation of the quantifications, monotone transformations can prevent overfitting. 
	
	We also estimated a GAM on this data set. In this analysis we fitted a nonmonotonic spline transformation for \emph{Tumor size} and all other predictors were defined with dummy variables. With these settings, GAM analysis closely resembles the nonmonotonic GLM-OS approach. This resemblance is supported by the similarity of the cross-validation results (\autoref{GLM-OS_tab:CVresults_BreastCancer}).
	
	\vspace{\baselineskip}
	
	We also estimated two ordinary logistic regressions with linear relation assumptions. 
	In the first analysis, we put linearity constraints on the \emph{Tumor size} and included all the other variables as categorical data by defining dummy variables. In the second analysis, we put linearity constrains on all four ordinal predictors (i.e.\ on \emph{Degree of malignancy}, \emph{Inv-nodes}, \emph{Age}, and \emph{Tumor size}). The latter is the standard (and only available) approach used if researchers want to preserve the category ordering. The cross-validation results for these models are also shown in \autoref{GLM-OS_tab:CVresults_BreastCancer}.
	
	When comparing the two logistic regression models with linear relation assumptions, the cross-validation results show that the prediction error for the dataset at hand (Apparent PE) is much larger when the linearity restrictions are put on all four variables compared to only on \emph{Tumor size}. However, the cross-validation shows that the expected prediction error is almost similar, although the misclassification rate is slightly higher for the model with most restrictions. 
	
	When comparing the results from ordinary logistic regression to the GLM-OS results, we see that the prediction accuracy (as measured by EPE) is in between the results of the models with nonmonotone and monotone scaling levels. This results suggests that applying no equal interval (=linear) restrictions on the ordering will give the worst predictions. The prediction error can be improved by imposing the linearity restrictions for only one or four predictors. However, the most beneficial option is to impose monotonicity instead of linearity. 
	
	Concluding, analyzing this data with mixed scaling levels that are appropriate for the mixed measurement levels of the predictors helped improve the prediction accuracy. When exploring the most suitable scaling levels, a cross-validation study is helpful to signal overfitting.

	\FloatBarrier 
	
	
	\section{Discussion}
	
	In this paper we have shown how OS transformations can be integrated in GLMs to transform predictors to optimize model fit and prediction accuracy. OS allows for nonlinear transformations that can be either nonmonotonic or monotonic, and are fit with a step or a spline function. 
	
	Transformations of the predictor variables have been integrated in GLMs before \citep{Book:GeneralizedAdditiveModels_HastiTib}. However, the OS methodology has several benefits compared to other methods.
	
	The strong focus of OS on categories, that is, regarding distinct values of variables as categories and thereby enabling handling of all variables as categorical no matter their measurement level, results in a more flexible analysis method.
	 Also, by not using dummy variables, interpretation of category results is easier. 
	 
	 While models like ordinary GLMs and GAMS use dummy variables to include categorical predictors, OS can quantify all categories directly because all dummies can be analyzed, as the OS-algoritm is not bothered by the perfect multicollinearity when dummies for all cat of a predictor are analyzed. Ordinary analysis can not handle perfect multicollinearity other than by excluding one of the dummies, so always requires a reference category which makes interpretation more difficult.
	 
	The quantifications are plotted against the original categories to visualize the transformations and thereby aid interpretation. Quantifications and model coefficients are also provided numerically.
	Hence, while ordinary GLMs focus on the fitted numerical results only, OS puts emphasis on visualizing the result.
	
	\vspace{\baselineskip}
	Another advantage of OS is the possibility to impose monotonicity restrictions on a transformation to preserve the ordering of categories. This monotonicity restriction can be beneficial in two ways. First of all, a monotone transformation makes interpretation easier since an increase in category implies an increase or decrease in the probability of the response. Furthermore, by imposing more restrictions on the transformation, there is a smaller risk of overfitting on the data at hand, which may reduce the prediction error for new data. 
	While models like ordinary GLMs and GAMS can only impose more restrictions by using a linear transformation, this might result in underfitting when relations are not linear. 
	Often monotonically transformed continuous variables show linearity for a sub-range of its values but is at a level thereafter or before, as is seen for the last predictor in \autoref{GLM-OS_fig:MonotoneAndNonMonotoneQuantifications_CMCdata}, or linearity within sub-ranges with different slopes, as seen for \emph{Age} in \autoref{GLM-OS_fig:MonotoneAndNonMonotoneQuantifications_CMCdata}. 
	The monotonic scaling levels offered by OS allow for transformations that are in between non-monotonic and linear, thereby enabling any kind of relation to be modeled.

	\vspace{\baselineskip}
	In a GLM-OS, the scaling level can be individually chosen for each predictor variable in the model. Hence, the most appropriate combination of transformation restrictions can be selected for each individual predictor. GLM-OS with mixed scaling levels is a provides a flexible analysis method that can be applied to a large variety of data types, ranging from unordered categorical data to (ordered) numerical data.

	\vspace{\baselineskip}
	Another feature of the OS technique is its group treatment of the levels of a categorical predictor variable. Namely, in OS a regression coefficient is obtained for each predictor to indicate its overall effect on the outcome (as in linear logistic regression) while such a diagnostic must be derived by the user from the results for logistic regression with dummy variables. In other words, in the OS setting the categories are not only analyzed individually, but together as a group, while in the dummy approach the grouping aspect is ignored.
	
	This handling of categories as a group is extremely useful when applying regularization techniques. Namely, in an OS analysis, regularization can be done directly on the regression coefficients since these are estimated separately from the quantifications. Three regularization methods, Ridge regression, the Lasso, and the Elastic Net, were already implemented in OS-regression (Regularized Optimal Scaling Regression; ROS Regression \citep{Art:MeulmanVanDerKooijROSS}), and these techniques can be implemented in GLM-OS in a similar manner. 
Alternatives like Group Lasso \citep{Art:GroupLasso} and Blockwise Sparse Regression \citep{Art:BlockwiseSparseRegression}, to regularize a group or block of instead of the individual variables, have been suggested to remedy this. However, applying regularization directly to the regression coefficients in the OS model is more straightforward and gives the same model fit. 
	Hence, the incorporation of regularization techniques is a useful future extension of GLM-OS.

	\vspace{\baselineskip}
	\section*{Acknowledgements}
	Clinicians and researchers involved in the the DINSTAP-project are gratefully acknowledged for providing the dataset used in this chapter (J.\,von Wietersheim, H.\,Weiss, I.\,Sammet, E.\,Gaus, E.\,Semm, D.\,Hartms, A.\,Eisenberg, R.\,Rahm, and J.\,K\"uchemhoff).
	
	\bibliographystyle{plainnat}
	\bibliography{ResearchBibliography.bib}
	
	\newpage
	\appendix
	\section{Supplementary material}

	\subsection{Calculating the gradient and Hessian of the negative log-likelihood function of the logistic regression model}
	\label{OSforGLM_app:LikelihoodLogReg_withDerivatives} 
	
	In a logistic regression function, the outcome is binary, i.e.\ $Y \in \{0,1\}$. The probability $\pi_i$ of having outcome $y_i=1$, given observed predictor variables $\xB_i$, is modeled. To avoid that the probability estimates are negative or exceed one, a logit link function maps the linear combination of predictor variables, $\eta_i = \xB_i \betaB$, onto the unit interval, i.e.\
	\begin{equation}
	\label{GLM-OS_Appendix_eq:LinkFuctionLogReg}
	P(y_i = 1) = \pi_i = \frac{1}{1 + \exp(-\eta_i)} = \frac{\exp(\eta_i)}{1 + \exp(\eta_i)} .
	\end{equation}
	Using this representation, the probability distribution for $Y_i$ is 
	\begin{displaymath}
	p(y_i) = P(Y_i = y_i) = \pi_i^{y_i} (1-\pi_i)^{1-y_i}.
	\end{displaymath}
	
	Since observations are assumed to be independent, the likelihood function is product of marginal probabilities, i.e.\
	\begin{align*}
	L(\etaB) &= \prod_{i=1}^n \pi_i^{y_i} (1-\pi_i)^{1-y_i}\\
	& = \prod_{i=1}^n \left(\frac{\pi_i}{1-\pi_i}\right)^{y_i}(1-\pi_i)\\
	& = \prod_{i=1}^n \exp(\eta_i)^{y_i}  \left[ 1-\frac{\exp(\eta_i)}{1 + \exp(\eta_i)}\right]\\
	& = \prod_{i=1}^n \exp(\eta_i)^{y_i}  \left[ \frac{1 + \exp(\eta_i) - \exp(\eta_i)}{1 + \exp(\eta_i)}\right]\\
	& = \prod_{i=1}^n \exp(\eta_i)^{y_i}   \frac{1}{1 + \exp(\eta_i)},\\
	\end{align*}
	and the corresponding log-likelihood function is
	\begin{align*}
	l(\boldsymbol{\etaB}) & = \log\left[ \prod_{i=1}^n \exp(\eta_i)^{y_i}   \frac{1}{1 + \exp(\eta_i)}\right]\\
	& = \sum_{i=1}^n y_i\eta_i -  \sum_{i=1}^n \log[ 1+  \exp(\eta_i)]. \\
	\end{align*}
	To simplify computations the negative log-likelihood 
	\begin{align*}
	l^{\text{-}}(\boldsymbol{\etaB}) & =  \sum_{i=1}^n \log[ 1+  \exp(\eta_i)] - \sum_{i=1}^n y_i\eta_i
	\end{align*}
	is minimized.
	The gradient of $l^{\text{-}}$ is the vector with elements
	\begin{align*}
	\frac{\partial l^{\text{-}}(\etaB)}{\partial \eta_i} & =[\log(1 + \exp(\eta_i ))]' -  [y_i \eta_i]'\\
	& =\frac{1}{1 + \exp(\eta_i )}[\exp(\eta_i )]' -  y_i  \\
	& = \frac{1}{1 + \exp(\eta_i )}\exp(\eta_i ) - y_i \\
	& = \frac{\exp(\eta_i )}{1 + \exp(\eta_i )} - y_i\\
	& = \pi_i - y_i.\\
	\end{align*}
	
	Since these partial derivatives are independent of $\eta_j$ for $j \neq i$ all second-order mixed partial derivatives are zero. Hence, the Hessian is a diagonal matrix with diagonal elements
	\begin{align*}
	\frac{\partial^2 l^{\text{-}}(\etaB)}{\partial \eta_i^2} 
	& = \frac{ [\exp(\eta_i)]' (1 + \exp(\eta_i)) - \exp(\eta_i)[1 + \exp(\eta_i)]'}{(1 + \exp(\eta_i))^2} - 0\\
	& = \frac{ \exp(\eta_i) (1 + \exp(\eta_i)) - \exp(\eta_i)\exp(\eta_i)} {(1 + \exp(\eta_i))^2} \\
	& = \frac{ \exp(\eta_i)} {1 + \exp(\eta_i)} - \frac{\exp(\eta_i)^2} {(1 + \exp(\eta_i))^2} \\
	& = \frac{ \exp(\eta_i)} {1 + \exp(\eta_i)} \left(1 - \frac{\exp(\eta_i)} {1 + \exp(\eta_i)} \right) \\
	& = \pi_i (1 - \pi_i).
	\end{align*}
	
	In matrix notation,
	\begin{align*}
	\textstyle{ \boldsymbol{\nabla} (\etaB) } & = \boldsymbol{\pi} - \yB; \\
	\mathbf{H}(\etaB) &= \text{diag} \left\{\boldsymbol{\pi} (1 - \boldsymbol{\pi} ) \right\},\\
	\end{align*} 
	where $\boldsymbol{\pi} = (\pi_1, \ldots, \pi_n)$ as defined in \eqref{GLM-OS_Appendix_eq:LinkFuctionLogReg}.

\end{document}